\documentclass{PoS}
\usepackage{graphicx}
\newcommand{\newc}{\newcommand}
\newc{\be}{\begin{equation}}
\newc{\ee}{\end{equation}}
\newc{\bea}{\begin{eqnarray}}
\newc{\eea}{\end{eqnarray}}
\newc{\ol}{\overline}
\newc{\wt}{\widetilde}
\newc{\bs}{\boldsymbol}
\newc{\m}{\mathcal}
\newc{\ra}{\rightarrow}
\newc{\lra}{\leftrightarrow}
\newc{\ba}{\begin{eqnarray}}
\newc{\ea}{\end{eqnarray}}
\newc{\pa}{\partial}
\newc{\D}{\Delta}

\newc{\nn}{\nonumber}

\newc{\diag}[1]{\ensuremath{{\rm diag} \left( #1 \right)}}

\def\beq{\begin{equation}}
\def\eeq{\end{equation}}
\def\bea{\begin{eqnarray}}
\def\eea{\end{eqnarray}}

\usepackage{amsmath}
\usepackage{amsfonts}
\usepackage{amssymb}
\usepackage{mathptmx}
\usepackage{slashed}
\usepackage{latexsym}
\usepackage{graphicx}
\usepackage{color}
\usepackage{float}
\usepackage{multirow}
\usepackage{amsmath}
\usepackage{amsfonts}
\usepackage{amssymb}
\usepackage{mathptmx}
\usepackage{slashed}
\usepackage{latexsym}
\usepackage{graphicx}
\usepackage{color}
\usepackage{float}
\usepackage{multirow}

\title{New Physics Phenomena and F-theory  GUTs}
\ShortTitle{B Physics $\&$ FGUTS}

\author{\speaker{George K. Leontaris}\\
        Physics Department, University of  Ioannina, GR-45110 Ioannina, Greece\\
        E-mail: \email{leonta@uoi.gr}}

\abstract{In this presentation the new physics implications of the  $B$-meson decay anomalies, observed at LHCb,  
    	are discussed. 	In the first part of the talk a brief overview of the experimental status is presented. 
    	In the second part,	a class of  semi-local F-theory GUT models with  additional  neutral gauge 
    	bosons are   proposed  which are capable of accounting for the  anomalous $B$-decay ratios $R_{K}$ and $R_{K^*}$.}

\FullConference{Proceedings of the Corfu Summer Institute 2017 "Workshop on the Standard Model and Beyond",\\
		2-10 September 2017\\
		Corfu, Greece }
		
\begin{document}

\section{Introduction}

\begin{figure}[!bth]
	\centering
	\includegraphics[scale=0.65]{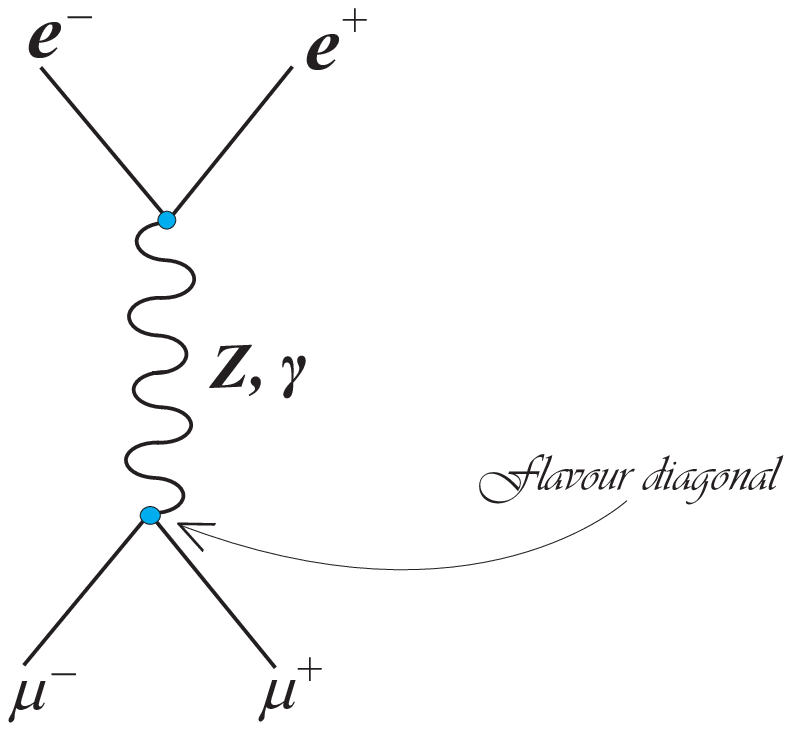}\;\;
	\includegraphics[scale=.65]{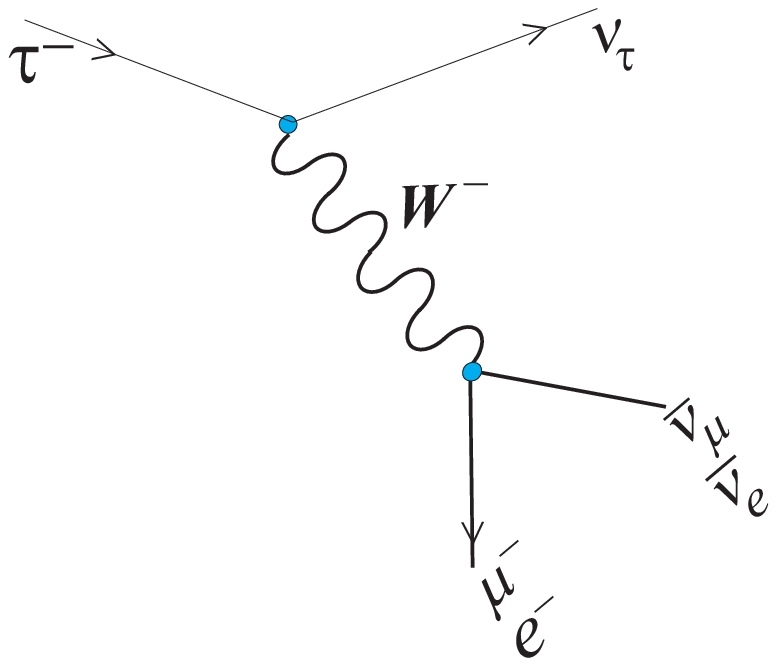}
	\caption{Representative interactions  involving leptons.  In the decay  $ { e^+e^- }\to { Z} \to {\mu^+\mu^-} $ the  $Z$ couplings are flavour diagonal in SM. In the second graph, the decay rates to the two lepton pairs $\mu\bar\nu_{\mu}$ and $e\bar{\nu_e}$ are the same. Experimentally it is found
		$	{ \Gamma(\tau^-\to \mu^-+\bar\nu_{\mu}+\nu_{\tau})\;{ =}\; \Gamma(\tau^-\to e^-+\bar\nu_{e}+\nu_{\tau})}$. }
	\label{f1}
\end{figure}

In the Standard Model of strong and electroweak interactions (SM) the couplings of the neutral  gauge boson $Z$  to lepton fields  are flavour independent. 
The $Z$-boson couples in the same way to all three families of lepton fields since fermions ($f$) with the same  charge $Q_f=T_3+Y$ have a universal coupling
$g^{Zff}= g\cos\theta_W { T_3}- g' \sin\theta_W { Y}$.  As a result, the tree-level  interactions conserve lepton flavour.  This property of the SM gauge 
interactions is usually called  Lepton Flavour Universality. The diagrams of two representative decays are shown in figure~\ref{f1}. In the left hand side,  
the diagram for the process $ { e^+e^- }\to { Z} \to {\mu^+\mu^-} $ is  shown where the $Z$-couplings to leptons are flavour diagonal. In the right hand side, 
the tree-level diagram for the decays $\tau\to \mu+\bar{\nu}_{\mu}+\nu_{\tau}$ and $\tau\to e+\bar{\nu}_{e}+\nu_{\tau}$ are drawn. Precision measurements show
 that the decay rates  are equal which imply equality of the corresponding weak coupling strengths.
Furthermore, in all experiments performed during the past decades,  precision tests regarding the decays of the $K$ mesons such as $K\to \ell\bar{\nu}_{\ell}$ where
$\ell=e,\mu$, are consistent with lepton universality. However, recent experimental data from LHCb regarding semileptonic decays of the
$B$-mesons~\cite{Aaij:2014ora,Aaij:2017vbb,MPA}, seem to be inconsistent with lepton universality. 
In the present talk we will focus on some of these  processes,  and in particular, those flavour changing  decays involving quark fields such as $b\to s\ell^+\ell^-$ where $\ell$ stand  for $e, \mu, \tau$ leptons. Suitable candidates for 
such decays  are the $B$ mesons composed of a quark-antiquark pair such as $B^+=\bar bu$ and $B^0=\bar b d$ with decay modes
\ba 
B^+&\to& K^+\ell^+\ell^-\label{BKL}\\
B^0&\to& {K}^{*0}\ell^+\ell^-~,\label{B0K0L}
\ea
where $K^+=\bar s u$ and  the spin-1 ${K}^{*0}$ is an excited state which subsequently  decays to 
an ordinary ground-state Kaon  and a pion,  ${K}^{*0}\to K^+\pi^-$. 
Once we have analysed the present experimental data, we will propose  solutions to this problem
in the framework of some theoretical models inspired from F-theory.

\section{Experimental Evidence}

We start  with a few facts regarding the theoretical results obtained in the context of the Standard Model (SM),
and the recent  experimental evidence which is in tension with these SM predictions.

\begin{figure}[!bth]
	\centering
	\includegraphics[scale=0.55]{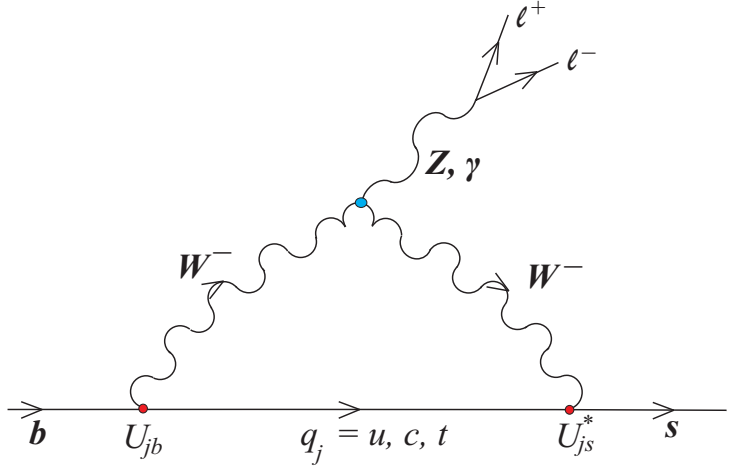}
	\caption{One-loop diagram leading to $b\to s$ flavour violation.
		Another contribution comes form an analogous box diagram.}
	\label{bsg}
\end{figure}

As is well known, experimentally we observe a suppression of the neutral currents compared to the charged ones involving lepton fields. 
For example,  the 
branching ratios of the following two semilectonic decays of the $B$ mesons are
(for related reviews see for example ~\cite{Altmannshofer:2014rta,Archilli,Grossman:2017thq})
\ba 
{\rm Br}(B\to D^0 \ell\bar{\nu})&=& 2.3\%,\\
{\rm Br}(B\to K^* \ell^+{\ell^-})&=& 5\times 10^{-7}~,
\ea 
where $D^0=c\bar u$ and $K^*=K\pi$.
The first  decay proceeds with a charged intermediate gauge boson at a much larger rate
compared to the second one which is mediated by neutral bosons. Focusing on the 
second case, we note that this is a  flavour changing  decay  in the quark sector, 
$b\to s\ell^+\ell^-$ (where $\ell$ stand  for $e, \mu, \tau$ leptons).
In the Standard Model these flavour changing  decays of the hadronic sector
($b\to s$) proceed through one-loop graphs 
and are suppressed by the CKM matrix elements (see figure~\ref{bsg}). 
Regarding the vertex $Z\,\ell^+\ell^-$, because of the universal nature  of the gauge couplings to leptons 
in the SM,  both  partners of the lepton pair $\ell^+\ell^-$ in a given reaction are  always members of  the same fermion 
family. Moreover, the SM predictions for  the branching ratios are the same
for all pairs  $e^+e^-,\, \mu^+\mu^-$ and  $\tau^+\tau^-$. Hence, the SM prediction  for 
all lepton pairs for the ratio of their branching ratios is expected to be equal to unity
\[R_{{X}_{ij}}=\frac{{\rm BR}(B\to X^+\ell_i^+\ell_i^-)}{{\rm BR}(B\to X^+\ell_j^+\ell_j^-)}
\approx 1,\; i,j=e,\mu, \tau;\;  X=K^+, K^0,\cdots \]
up to insignificant corrections. The experimental 
determination of the ratio $R_X$ is preferable  because all
theoretical uncertainties stemming from the hadronic part in the branching ratios cancel out  in the ratios. 

\begin{figure}[!bth]
	\centering
	\includegraphics[scale=0.6]{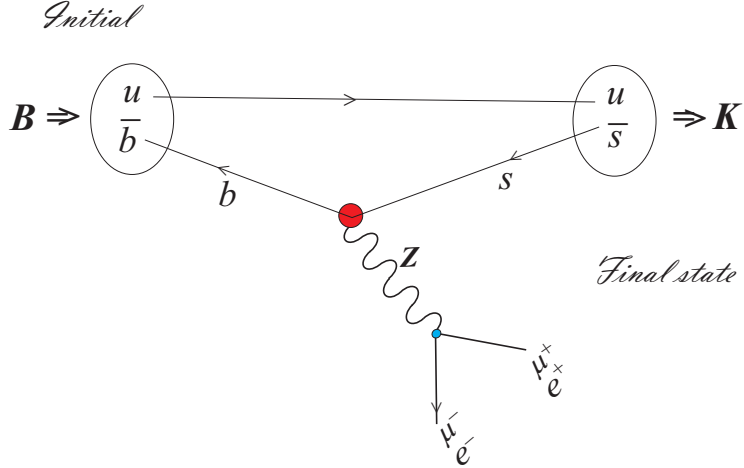}
	\caption{An illustration of the $B\to K\ell_i^+\ell_i^-$ decay in SM.  The red bullet stands for the one-loop
		contribution in figure~\ref{bsg}. }
\end{figure}

Such  decays, however,  are very sensitive to contributions from interactions violating lepton flavour universality.  
In many extensions of the Standard Model new particles are predicted (such as neutral gauge bosons and leptoquarks)
which can enhance or reduce the  decay rates and modify the angular distribution of the products of the above processes.  
The decay~(\ref{B0K0L}), in particular, involves four particles in the final state and allows for a precise angular 
reconstruction in several observables.  In the muon channel for example,  the experiment can 
measure  the polarization of $K^*$ and the dimuon pair, as well as the invariant  mass square $q^2$, the angle $(\mu, K)$, etc.

Interestingly, recent reports~\cite{Aaij:2017vbb} from the LHC experiments on the branching ratios of  B-decays to strange hadrons and lepton pairs
have displayed deviations form lepton universality.  The ratio of the branching ratios  
$B\to K^+ \ell\bar \ell$ for $\ell =\mu$ or $e$, measured in the LHCb experiment is
\[
R_{K}=\frac{{\cal B}(B\to K^+\mu^+\mu^-)}{{\cal B}(B\to K^+e^+e^-)}
= 0.745\pm 0.09(stat)\pm 0.036(syst)~,
\]
where ``(stat)'' , ``(syst)'' indicate  statistical and  systematic uncertainties respectively. 
This measurement is 2.6 standard deviations below the SM prediction. 
The decay rate of this reaction is integrated over the range of the squared dilpeton invariant mass $q^2$ and
the region is taken to be 
$1$GeV$^2 < q^2< 6$ GeV$^2$, which is away from the resonance region $B^+\to J/\psi (\ell^+\ell^-) K^+$,
in order to have a clear experimental signature to compare with the theoretical predictions.

Similarly, the experimental result on the ratio of branching ratios $B\to K^* \ell\bar \ell$ 
is given by 
\[R_{K^*}=\frac{{\cal B}(B\to K^*\mu^+\mu^-)}{{\cal B}(B\to K^*e^+e^-)}\approx\left\{\begin{array}{cc}
0.660\begin{array}{c}{ +0.110}\\
{ -0.070}\\
\end{array}\pm 0.024
\\
0.685\begin{array}{c}{ +0.113}\\
{ -0.069}\\
\end{array}\pm 0.047
\end{array}
\right.~,
\]
for the ranges $(2m_{\mu})^2< q^2< 1.1$ GeV$^2$ and   $1.1$GeV$^2 < q^2< 6$ GeV$^2$  respectively.  
 The final products $K^+,\pi^-$  associated with the $K^{0*}$ meson 
 are those with an invariant mass  within $\sim 100$ MeV  of the value $m_K^*=892 {\rm MeV}$.
  (These data correspond to an integrated luminocity of $3 fb^{-1}$ of proton-proton collisions by LHCb during 2011-2012).

\begin{figure}[!bth]
	\centering
	\includegraphics[scale=0.75]{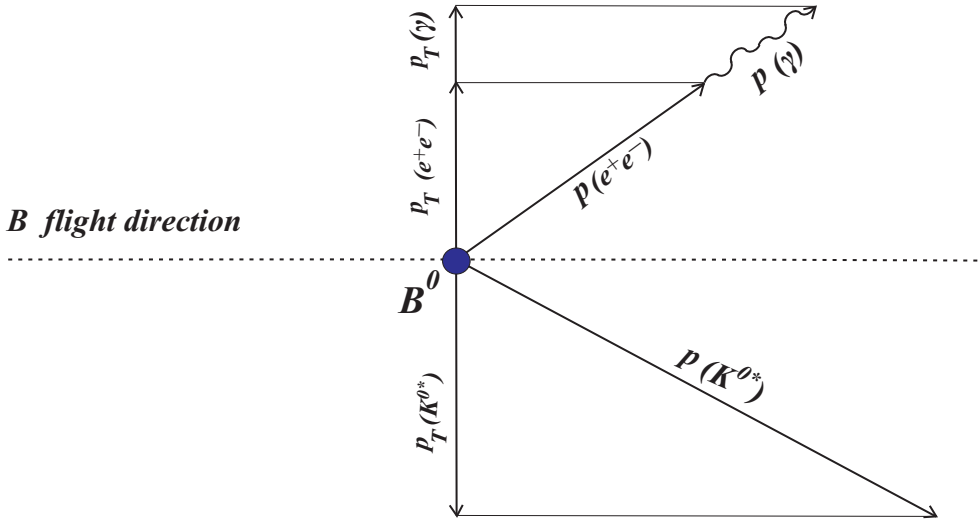}
	\caption{The topology of $B\to K^*e^+e^-$ decay and  Bremsstralung effect recovery in accordance to~\cite{Aaij:2017vbb}}
	\label{brem0}
\end{figure}

However, as already noted, according to the universality of gauge interactions in the leptonic sector, the SM theoretical predictions
are  $R_K({SM})=R_K^*(SM)\approx 1$ up to insignificant electromagnetic corrections of ${\cal O}(m_{\mu}/M_B)$.
Thus, the measurments of  both ratios indicate a deficit in the same direction. 
 Experimentally, in LHCb there are differences regarding the treatment of the decays 
to a $\mu^+\mu^-$ or $e^+e^-$ final state since these pairs behave differently in their flight through the material of the detector. 
This is because electrons, being lighter than muons,  emit much bremsstralung and as a result  there is significant reduction of 
the momentum, thus the efficiency in the muon channel is much better that in the electron's. 
This defficiency, however,  is improved by a recovery procedure based on the evaluation of the
difference between the $p_T$ of the $K^{*0}$ meson and that of the $e^+e^-$ pair,
where both are calculated with respect to the direction of the flight of the  $B^0$ meson (see fig.\ref{brem0}).

Eventually, assuming that this experimental evidence remains valid in future analysis,  these  defficits in the muon channel are unexplained 
in the Standard Model framework and  new physics interactions are required, in order to discriminate the electron from muon ratios
in the semileptonic $B$ meson decays.  Indeed, it is commonly believed that  the Standard Model is not the final theory of fundamental interactions, 
but just an effective low energy limit of some  grand unified theory (GUT), possibly embedded in a string scenario.  
In general, any GUT which includes the SM gauge group predicts new physics phenomena and deviations from the SM predictions. 
Some common characteristics of a wide class of models derived in  such a framework are  exotic colour tiplets and singlet fields, 
new gauge bosons $Z'$ associated with additional abelian symmetries etc~\footnote{There is a vast number of related papers in the literature. 
For an incomplete list see~\cite{AllPapers}.}. All these 
new ingredients can in principle mediate new exotic processes or enhance 
others already observed in recent experiments. Since  the experimental results of anomalous $B$ meson decays 
remain  in place through the last few years, it is worth exploring viable SM  extentions  to explain them.

\begin{figure}[!bth]
	\centering
	\includegraphics[scale=0.65]{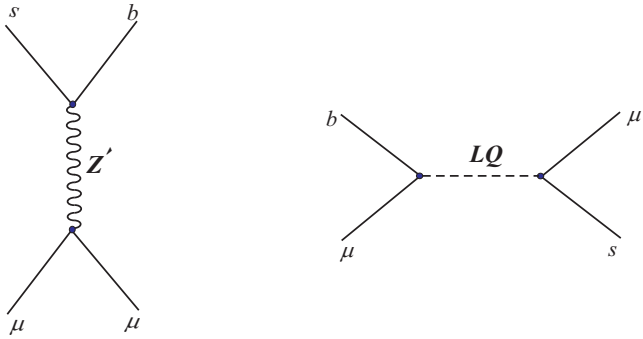}
	\caption{ $Z'$ boson and leptoquark contributions to  $b\to s\mu^+\mu^-$ }
	\label{ZpLQ}
\end{figure}

The most popular scenarios to interpret  the aforementioned deviations   include either 
exotic  leptoquarks,  or  additional neutral gauge bosons which couple differently
to the three fermion families.
Given the fact that the composition of mesons in $B\to K$ decays is $B=u\bar b$ and $K=u\bar s$,
at the fundamental level these results are associated with the $ b\to  s\mu^+\mu^-$ transition.  The tree-level graphs for such processes are shown in figure~\ref{ZpLQ}.

\section{The effective Hamiltonian description}

The contributions from  SM physics as well as from  any theory  beyond the SM, can be
described by an effective theory where the heavy degrees of freedom have
been integrated out. The relevant Hamiltonian is  parametrised in terms of the Wilson coefficients $C_k$, which display the strength of the interaction, 
and a number of quantum operators ${\cal O}_k$ which encode the Lorentz structure~\cite{Altmannshofer:2014rta}:
\be 
{\cal H}_{eff} = -\frac{4 G_F}{\sqrt{2}}V_{tb}V_{ts}^*\frac{\alpha}{4\pi} \sum_k\left( C_k(\mu) {\cal O}_k(\mu)+  C'_k(\mu) {\cal O}'_k(\mu)\right)~\cdot
\ee 
In the above formula the following quantities are involved:  
$G_F=\sqrt{2}/(4v^2)$ (where $v=174$ GeV) is the Fermi constant,  $V_{tb}\approx 0.95, V_{ts}^*\approx 0.4$  
are CKM elements relevant to the specific transition, and $\alpha\approx 1/128$ is the fine structure constant computed  at the scale $m_b$.
 The Wilson coefficients $ C _k(\mu)$  and the dimension-six quantum operators  ${\cal O}_k(\mu)$ are defined at the scale $\mu=m_b$.
The Lorenz invariant operators relevant to the processes are 
\ba 
{\cal O}_9={\bar s}\gamma_{\lambda} { P}_L b \bar{\ell} \gamma^{\lambda}\ell
&&
{\cal O}'_9={\bar s}\gamma_{\lambda} { P}_R b \bar{\ell} \gamma^{\lambda}\ell\label{O9}
\\
{\cal O}_{10}={\bar s}\gamma_{\lambda} { P}_L b \bar{\ell} \gamma^{\lambda}\gamma_5\ell
&&
{\cal O}'_{10}={\bar s}\gamma_{\lambda} { P}_R b \bar{\ell} \gamma^{\lambda}\gamma_5\ell\label{O10}
\ea
where $P_L=\frac 12(1-\gamma_5)$ and $P_R=\frac 12(1+\gamma_5)$.

\begin{figure}[h]
	\centering
	\includegraphics[scale=.75]{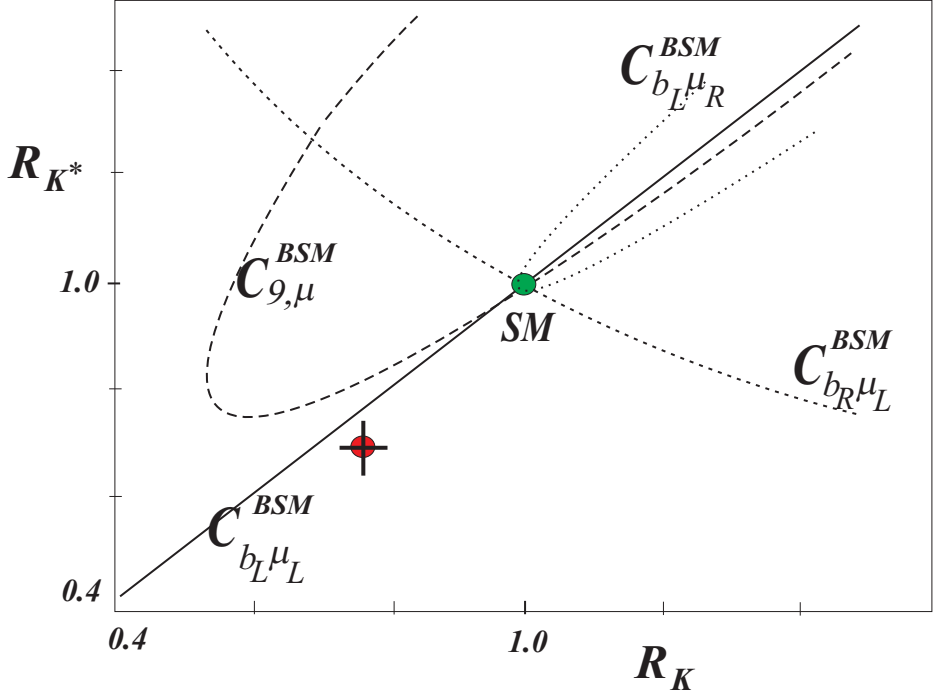}
	\caption{  Correlations of {$R_K/R_{K^*}$} for various Lorentz operators. The 
		green dot represents the SM prediction and the red one the experimental value.  }
	\label{Cor}
\end{figure}

There are several theoretical uncertainties in the computation, the most important come  from the form factors
and the contributions of the hadronic weak Hamiltonian which emerges from the assumed factorisation of the
amplitude into a hadronic and a leptonic part. These are discussed for example in~\cite{Altmannshofer:2014rta}.
Focusing in $B^+\to K^+\mu^+\mu^-$ in the limit of
vanishing lepton mass, the decay rate can be written as
\be 
\frac{d\Gamma}{d q^2}= \frac{G_F^2\alpha^2}{(4\pi)^5 m_B^3} \left|V_{tb}V_{bs}^*\right| g^{3/2}(m_B,m_{K^*},q^2) (|F_V|^2+|F_A|^2)
\label{Gq2}
\ee 
The  quantities in (\ref{Gq2}) are~\cite{Altmannshofer:2014rta}: \[g(x_i)=\sum_{i=1}^3x_i^2-2 x_1 x_2-2 x_2 x_3-2 x_1 x_3,\;
 F_V=(C_9+C_9')f_+(q^2),\; F_A=(C_{10}+C_{10}')f_+(q^2)+h_K
\]
where $f_+(q^2)$ is a QCD factor, $h_K$ non-factorisable contributions of ${\cal H}_{eff}$, while $C_{7},C_{7}'$ contributions are ignored. 
As can be seen, there are several operators with different Lorentz structure 
that could be present in the effective Hamiltonian, but only a few of them
are related to the present LHCb data.  
The correlations of {$R_K/R_{K^*}$} deviations  and the corresponding chiral operators generated by { New Physics} in the {$\mu$} sector are plotted in ref~\cite{DAmico:2017mtc} and are roughly depicted  here in figure~\ref{Cor}.
The LHCb data can be interpreted assuming a negative  contribution to the Wilson  coefficient $C_9$,
with best fit value $C_9^{NP}\approx -1.1$.  The case of  $C_9^{NP}=-C_{10}^{NP}\approx -0,5$, which is
 an $SU(2)_L$ invariant solution is also possible.  Both solutions 
ar compatible with negligible values of  the Wilson coefficients $C_9', C_{10}'$.

Let's discuss the  case where the data are fit with the first solution. 
It is, in principle, possible that the required operator is  generated 
from a tree-level process mediated by a TeV scale $Z'$ boson that couples
 to the $b,s$ quarks and  the leptons.  
 This neutral gauge boson can be associated with a
new $U(1)'$ gauge symmetry spontaneously broken at the TeV scale.
$Z'$ couples to a neutral current $L \supset Z^{'\lambda}J_{\lambda}^{'0}$
where the couplings to the third quark generation differ from those to the 
first and second ones. To give an estimate, we use the toy model presented 
in ref~\cite{Allanach:2015gkd}, where the current is assumed to be of the form:
\ba 
J_{\lambda}^{'0}&=& g_{tL}(\bar b \gamma_{\lambda} P_L b+\bar t \gamma_{\lambda} P_L t)+
g_{\mu} (\bar \mu\gamma_{\lambda}\mu+\bar \nu_{\mu}\gamma_{\lambda}\nu_{\mu})
\nn\\
&&+ g_{qL}\sum_{q=u,d,s,c}(\bar q\gamma_{\lambda}P_Lq)+(g_{tL}-g_{qL})V_{ts}^*V_{tb}\bar s\gamma_{\lambda}P_Lb+h.c.
\ea 

Then, the new physics contribution to the Wilson coefficient $C_9$ is~\cite{Allanach:2015gkd}
\be 
C_9^{NP}= -\frac{\pi g_{\mu} (g_{tL}-g_{qL})}{2\sqrt{2} G_F M_{Z'}^2 \alpha \cos^2\theta_W}
\approx  -\frac{\pi { g_{\mu} (g_{tL}-g_{qL})}}{ c_W^2 \left(\frac{ M_{Z'}}{{ 2}\,{\rm TeV}}\right)^2 }
\ee
\begin{figure}[!bth]
	\centering
	\includegraphics[scale=0.75]{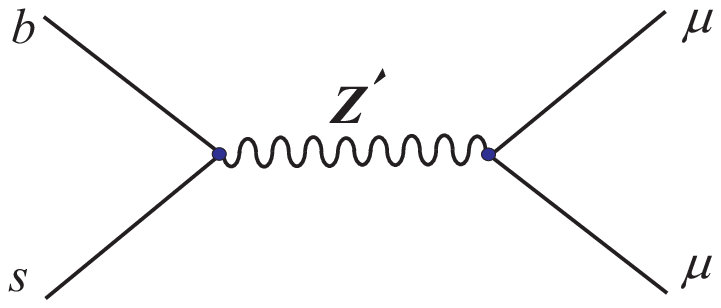}
	\caption{$Z'$ boson and $b\to s\mu\mu$ decay}
	\label{brem}
\end{figure}
For couplings  $g_{\alpha}\sim \frac 12$, and $M_{Z'}\sim $ few TeV, ( which is consistent
with the renormalisation group analysis of models admitting such $U(1)$ symmetries\cite{Karozas:2017hog}), we can  obtain  $ C_9^{NP}\sim -1$
which reduces the total value of $ C_9$ by $\sim 25\%$ in accordance with the experimental observations.   Notice however, that any theory beyond the SM interpreting the above anomalies
must also respect other constraints too, and in particular, with the
stringent bounds~\cite{Buras:2015qea} of Kaon  decays such as
${\cal B}(K^+\to \pi^+ \nu\bar \nu)\approx 17\pm 10\times 10^{-11}$ and 
${\cal B}(K_L\to \pi^0 \nu\bar \nu)< 2.6\times 10^{-8}$.

\section{Non-universal $U(1)$s in local F-theory models}

A natural way to implement the idea of a new gauge boson $Z'$  is within an F-theory
 framework~\cite{Beasley:2008dc,Bouchard:2009bu}~\footnote{For 	model building with F-theory and early references 
	see reviews~\cite{Heckman:2010bq,Leontaris:2012mh,Maharana:2012tu}}. The so derived model(s) can explain the
date either through the existence of a new light neutral boson coupled 
differently to families~\cite{Romao:2017qnu}, and/or in the presence of a vectorlike family\cite{King:2017anf,Romao:2017qnu}.  
The interpretation of the data with leptoquarks is also  a viable possibility, since plenty of such states appear 
in a wide class of F-theory models. In this talk, we will focus only on the first case and 
present  a semi-local F-theory $SU(5)$ GUT augmented with a $U(1)'$ 
factor coupled non-universally  to leptons,
arising from a variant of $E_6$ models with ${\cal Z}_2$ monodromy~\cite{Callaghan:2011jj}.
We start with  the following  symmetry breaking,
\ba
E_8
& \supset & E_6\times U(1)'_\perp\times U(1)_\perp \nn \\
& \supset & SO(10)\times U(1)_{\psi}\times U(1)'_\perp\times U(1)_\perp \label{SO(10)} \\
& \supset & SU(5)_{GUT} \times  U(1)_{\chi}  \times U(1)_{\psi} \times  U(1)'_\perp\times U(1)_\perp ,
\ea
Imposing a ${\cal Z}_2$ monodromy,  the symmetry of the model reduces to   $SU(5)_{GUT} \times  U(1)^3$ and
at the same time a top Yukawa coupling is allowed at tree-level\cite{Bouchard:2009bu}.  
We choose the following basis for the Cartan generators corresponding to the three remaining abelian factors:
\ba 
Q_\perp=\frac{1}{2\sqrt{3}}{\rm diag}(1,1,-2,0,0),\;
Q_\psi=\frac{1}{2\sqrt{6}}{\rm diag}(1,1,1,-3,0),\;
Q_\chi=\frac{1}{2\sqrt{10}}{\rm diag}(1,1,1,1,-4)~\cdot\nn
\ea
Next, since we would like to explain the experimental results invoking the existence of a TeV scale  neural gauge boson $Z'$, we assume 
that a  low energy $U(1)^\prime$ is generated by a linear combination
of the unbroken $U(1)$'s: 
\be
Q^\prime = c_1 Q_\perp + c_2 Q_\psi + c_3 Q_\chi~\cdot \label{Qprime}
\ee
Furthermore, in order to retain $SU(5)_\perp$ normalisation, the coefficients $c_i$  are subject to the condition
\be  
c_1^2 + c_2^2 + c_3^2 =1 \label{cont}
\ee 
An effective $SU(5)$ model now can be constructed  by assuming suitable fluxes along the  
$U(1)$  factors. These fluxes generate chirality for the $10/\overline{10}$ and $5/\bar 5$ representations
residing in the intersections (matter curves) of  the $SU(5)$ GUT divisor and the 7-branes associated 
with the abelian factors. There are initially ten matter curves available to accommodate
the fiveplets and five matter curves for the tenplets but after the ${\cal Z}_2$ monodromy action
they reduce to seven and four respectively~\cite{Dudas:2009hu,King:2010mq}.
We designate the corresponding numbers of the $SU(5)$ multiplets on the matter curves with $M_{5_i}, M_{10_j}$
respectively.  Furthermore,  we may turn on  
flux along the hypercharge  generator $U(1)_Y$ which breaks $SU(5)$  down to SM and at the same time 
splits the $10, \overline{10}$ 
and $5,\bar 5$'s  into different numbers of Standard Model multiplets. Parametrising the hyperfluxes with the integers
 $N_{7,8,9}$   and assuming a linear combination  of them,  $N_y$, to be the   hyperflux piercing  a given  matter curve,
  the 10-plets and 5-plets  split according to the following pattern
\be
{10}_{{j}}=
\left\{\begin{array}{cl}{\rm
		SM\;\;field}&{\;\;\;\;\;\rm flux\, units }\\
	n_{{(3,2)}_{+\frac 16}}-n_{{(\bar 3,2)}_{-\frac 16}}&=\;M_{10_j}\\
	n_{{(\bar 3,1)}_{-\frac 23}}-n_{{(
			3,1)}_{+\frac 23}}&=\;M_{10_j}-N_{y_j}\\
	n_{(1,1)_{+1}}-n_{(1,1)_{-1}}& =\;M_{10_j}+N_{y_j}\\
\end{array}\right.\;,\;\;\;
\\
{5}_{{i}}=
\left\{\begin{array}{cl}{\rm
		SM\;\; field}&{\;\;\;\;\;\rm flux\, units }\\
	n_{(3,1)_{-\frac 13}}-n_{(\bar{3},1)_{+\frac 13}}&=\;M_{5_i}\\
	n_{(1,2)_{+\frac 12}}-n_{(1,2)_{-\frac 12}}& =\;M_{5_i}+N_{y_i}\\
\end{array}\right.\,\cdot
\ee

\begin{table}[H]
	\small
	\centerline{
		\begin{tabular}{|c|c|c|c|c|c|c|c|c|c|c|c|c|c|c|}
			\hline
			$c_1$                 & $c_2$                            & $c_3$                            & $M_{5_{H_u}}$ & $M_{5_1}$ & $M_{5_2}$ & $M_{5_3}$ & $M_{5_4}$ & $M_{5_5}$ & $M_{5_6}$ & $M_{10_t}$ & $M_{10_2}$ & $M_{10_3}$ & $M_{10_4}$ & $N_{7,8,9}$  \\ \hline
			$0$ & $\frac{\sqrt{\frac{15}{34}}}{2}$ & $\frac{11}{2 \sqrt{34}}$ & $0$ & $0$ & $0$ & $0$ & $-1$ & $-3$ & $1$ & $1$ & $2$ & $ -1$ & $1$ & $N_7=1$   \\
			$-\frac{\sqrt{\frac{5}{6}}}{2}$ & $-\frac{5 \sqrt{\frac{5}{3}}}{8}$ & $\frac{3}{8}$ & $0$ & $1$ & $-1$ & $0$ & $-1$ & $-2$ & $0$ & $2$ & $-1$ & $1$ & $1$ & $N_9=1$ \\
			$\frac{\sqrt{3}}{2}$ & $-\sqrt{\frac{3}{32}}$ & $\sqrt{\frac{5}{32}}$ & $0$ & $0$ & $0$ & $1$ & $-3$ & $-1$ & $0$ & $2$ & $1$ & $1$ & $-1$ &  $N_8=1$  \\\hline
		\end{tabular}
	}
	\caption{Flux parameters and $c_i$ coefficients for three models}
	\label{dataex1to33}
\end{table}
\begin{table}[h]
	\small
	\centering
	\begin{tabular}{c|c|c||c|c||c|c|}
		\cline{2-7}
		&             & \multicolumn{1}{c||}{Model A}                                      & \multicolumn{2}{c||}{Model B}                                      & \multicolumn{2}{c|}{Model C}                                 \\ \hline
		\multicolumn{1}{|c|}{Curve}                        & $Q^\prime\sqrt{85}$ & SM Content                                    & $Q^\prime\sqrt{10}$ & SM Content                                    & $Q^\prime$     & SM Content                                    \\ \hline
		\multicolumn{1}{|c|}{\multirow{2}{*}{$5_{H_u}$}}   & $-4$                & $H_u$                                         & $\frac{3}{2}$       & $H_u$                                         & $-\frac{1}{2}$ & $H_u$                                         \\
		\multicolumn{1}{|c|}{}                       & ---                 & ---                                           & ---                 & ---                                           & ---            & ---                                           \\
		\multicolumn{1}{|c|}{\multirow{2}{*}{$5_1$}}     & ---                 & ---                                           & $\frac{1}{4}$       & $\overline{d^c}$                              & ---            & ---                                           \\
		\multicolumn{1}{|c|}{}                           & $4$                 & $H_d$                                         & ---                 & ---                                           & $-\frac{1}{4}$ & $L$                                           \\
		\multicolumn{1}{|c|}{\multirow{2}{*}{$5_2$}}   & ---                 & ---                                           & ---                 & ---                                           & ---            & ---                                           \\
		\multicolumn{1}{|c|}{}                          & $\frac{3}{2}$       & $L$                                           & $1$                 & $d^c+2 L$                                     & $\frac{1}{2}$  & $H_d$                                         \\
		\multicolumn{1}{|c|}{\multirow{2}{*}{$5_3$}}     & ---                 & ---                                           & ---                 & ---                                           & $0$            & $\overline{d^c}$                              \\
		\multicolumn{1}{|c|}{}                            & $-\frac{7}{2}$      & $L$                                           & $-\frac{3}{2}$      & $H_d$                                         & ---            & ---                                           \\
		\multicolumn{1}{|c|}{\multirow{2}{*}{$5_4$}}    & ---                 & ---                                           & ---                 & ---                                           & ---            & ---                                           \\
		\multicolumn{1}{|c|}{}                    & $\frac{3}{2}$       & $d^c$                                         & $\frac{9}{4}$       & $d^c+L$                                       & $-\frac{1}{4}$ & $3 d^c + 2 L$                                 \\
		\multicolumn{1}{|c|}{\multirow{2}{*}{$5_5$}}    & ---                 & ---                                           & ---                 & ---                                           & ---            & ---                                           \\
		\multicolumn{1}{|c|}{}                          & $-\frac{7}{2}$      & $3 d^c+2L$                                    & $-\frac{1}{4}$      & $2 d^c + L$                                   & $-\frac{3}{4}$ & $ {d^c}+ L$                                   \\
		\multicolumn{1}{|c|}{\multirow{2}{*}{$5_6$}}     & $6$                 & $\overline{d^c}+\overline{L}$                 & $-1$                & $\overline{L}$                                & $0$            & $\overline{L}$                                \\
		\multicolumn{1}{|c|}{}                         & ---                 & ---                                           & ---                 & ---                                           & ---            & ---                                           \\ \hline
		\multicolumn{1}{|c|}{\multirow{2}{*}{$10_t$}}       & $2$                 & $Q+2 u^c$                                     & $-\frac{3}{4}$      & $2 Q+3 u^c+e^c$                               & $\frac{1}{4}$  & $2Q+3 u^c+e^c $                               \\
		\multicolumn{1}{|c|}{}                            & ---                 & ---                                           & ---                 & ---                                           & ---            & ---                                           \\
		\multicolumn{1}{|c|}{\multirow{2}{*}{$10_2$}}         & $2$                 & $2 Q+u^c+3 e^c$                               & ---                 & ---                                           & $-\frac{1}{2}$ & $Q + u^c+ e^c$                                \\
		\multicolumn{1}{|c|}{}                           & ---                 & ---                                           & $-\frac{1}{2}$      & $\overline{Q}+\overline{u^c}+ \overline{e^c}$ & ---            & ---                                           \\
		\multicolumn{1}{|c|}{\multirow{2}{*}{$10_3$}}         & ---                 & ---                                           & $\frac{7}{4}$       & $Q+u^c+e^c$                                   & $\frac{1}{4}$  & $Q+2 e^c$                                     \\
		\multicolumn{1}{|c|}{}                                & $\frac{1}{2}$       & $\overline{Q}+\overline{u^c}+ \overline{e^c}$ & ---                 & ---                                           & ---            & ---                                           \\
		\multicolumn{1}{|c|}{\multirow{2}{*}{$10_4$}}        & $\frac{11}{2}$      & $Q+u^c+e^c$                                   & $-\frac{3}{4}$      & $Q+2 e^c$                                     & ---            & ---                                           \\
		\multicolumn{1}{|c|}{}                            & ---                 & ---                                           & ---                 & ---                                           & $\frac{1}{4}$  & $\overline{Q}+\overline{u^c}+ \overline{e^c}$ \\ \hline
	\end{tabular}
	\caption{The low energy spectrum for the three models of  Table~\ref{dataex1to33}}. (There are also singlet fields~\cite{Romao:2017qnu} not shown in this table)
	\label{ex1to33}
\end{table}
The integers $M_{10_j}, M_{5_i}$ representing the multiplicities and the coefficients  $c_k$ defining the
linear combination~(\ref{Qprime}) are subject to anomaly cancellation conditions~\cite{Romao:2017qnu} (see also related
work for $B$-decays~\cite{Ellis:2017nrp}) and the
constraint~(\ref{cont}) respectively. There are numerous solutions~\cite{Romao:2017qnu} to these constraints.
Three examples  are given in  Table~\ref{dataex1to33}  where the flux data and solutions for the coefficients $c_i$ are presented.
The spectrum of these three models is determined by the integers 
$M_i, N_j$ which are computed in consistency with the anomaly cancellation conditions and
can be seen in  Table~\ref{ex1to33}. We also mention the existence of 
several singlet fields in the spectrum which may acquire non-zero vevs and,
amongst other things, they play a crucial r\^ole in  the anomaly cancellation conditions and the
Yukawa couplings of the effective theory~\cite{Romao:2017qnu}.

From Table~\ref{ex1to33}  we see that model $C$ has 3 chiral families and a vectorlike one. 
A model consistent with all the experimental constraints requires a careful  distribution
of the chiral particle content on the various matter curves.  Thus, in view of the stringent  experimental constraints
from the  $K^0-\bar K^0$  oscillations, the three families are accommodated so that their 
left-handed quark doublets  have equal charges, $Q'=\frac 14$, with respect to  $U(1)'$. 
Also, the  three down quarks and lepton doublets are distributed in `curves' of fiveplets with the same charge 
$Q'=-\frac 14$,  in Table~\ref{ex1to33}.  We may accommodate the  vectorlike pair of lepton 
doublets  on the curves $5_{5,6}$ which have non-universal $U(1)'$ 
charges and can mix differently with the lepton doublets  $L_{1,2}$, inducing non-universal couplings 
in the physical left-handed electrons and muons. This  can account for the observed ratios $R_{K}$ and $R_{K^*}$
in $B$ decays  and it is discussed in detail in~\cite{King:2017anf} and ~\cite{Romao:2017qnu}.

In conclusion, we have presented a class of models with   $SU(5)\times U(1)'$ gauge symmetry embedded in $E_8\supset
 SU(5)\times U(1)_{\perp}^4$. The abelian factor $ U(1)'$ is a linear combination of the three 
remaining $U(1)_{\perp}$  factors, after a monodromy is implemented and the corresponding gauge boson $Z'$  displays 
non-universal gauge couplings  to fermion families.
From class $C$ of Table~\ref{ex1to33}, a  phenomenologically promising model  emerges with universal couplings to
 the three chiral fermion generations and only one 
single vectorlike family having non-universal couplings. Their couplings modify   the branching ratios of the 
$B$-decays in accordance with the recently observed $\mu^+\mu^-$ deficit.

\vspace*{.5cm}
\noindent
{\it The author would like to thank the organisers of the ``Corfu Summer Institute'' for kind hospitality}

\end{document}